\documentclass[11pt,a4paper]{article}
\usepackage{main}
\usepackage{times}
\usepackage{latexsym}
\usepackage{listings}
\usepackage{algorithm}
\usepackage{algpseudocode}
\usepackage{url}
\usepackage{amsmath}
\usepackage{mathtools}

\aclfinalcopy 
    
\title{Honors Thesis: Transformer-based Program Synthesis for Low-Data Environments}

\author{Jack Roper}

\date{May 2022}

\begin{document}
\maketitle
\begin{abstract}
  Recent advancements in large pre-trained transformer models (GPT2/3, T5) have found use in program synthesis to generate programs that satisfy a set of input/output examples. However, these models perform poorly on long-horizon and low-data tasks, and often don't seem to understand the semantics of the languages they generate. We investigate an approach that tackles both of these issues, by using attributed context-free-grammars of programming languages to generate programs, and then analyzing generated programs so that they can be annotated with compile and runtime attributes, such as types, so that information about the program can be remembered during long-horizon generation. We firstly find that synthesized datasets can be made efficiently and can provide transformer models with enough data in order to perform well on some synthesis tasks. We also find that giving models access to program attributes is especially effective in low-data environments, and tends improve the quality and reduce errors of transformer-generated programs. 
\end{abstract}

\section{Introduction}

A classic task in program synthesis is, given a specification, to generate a program in some given language that fits this specification. Specifications can be given as logical constraints, natural language descriptions, or other formats that describe a program. Classically, such synthesis is done via. enumerative or SMT-based methods \cite{DBLP:conf/cav/BarrettCDHJKRT11}, that usually generate a symbolic representation of the desired program. However, these methods suffer serious drawbacks: The main among them being how the search space of programs tends to blow up exponentially with respect to the possible length of the program. Pruning the search space can be done using type data \cite{l2} or other compile-time information, but the problem of the exponentially-growing search space still stands, and prohibits long-horizon program generation.

Attention has recently instead shifted towards using neural-network-based methods to generate programs. One such neural network architecture is the transformer, a neural network which takes in as input a string known as the "prompt", and then outputs the prompt appended with an additional string \cite{transformerOrig}. This can be seen as a sort of "autocomplete" task, where the transformer predicts how to finish a sentence or autocomplete a string. Many tasks in program synthesis can be thought of as a transformer task. For example, a prompt can be a function header or definition, and we may expect the transformer to return a valid function that matches that header. Another way we can use the transformer is to prompt it with input/output examples of a program, and then to expect the transformer to finish the prompt with a program that matches those examples.

Transformer-based synthesis allows for program synthesis given many different kinds of specifications. And, while transformer-based methods are computationally expensive, synthesis time grows only linearly with the size of the generated program, as opposed to classical synthesis methods, which tend to take exponential time with respect to the depth of the program. Currently, multiple large transformer models have already been created and tested on program synthesis. Models such as OpenAI's Codex \shortcite{chen2021evaluating} or GPT-Neo \shortcite{gpt-neo} have been pre-trained on a large corpus of source code, and can be used to generate programs with decent results. However, there are large, glaring problems with these models.  

The largest issue with these models is the amount of training data required. For example, Codex was trained on over 159 GB of available open source code \cite{chen2021evaluating} from over 164 thousand Python Github Repos, while GPT-Neo \shortcite{gpt-neo} was trained on a ~825GB cross-language dataset called The Pile \cite{thepile}. For languages with small amounts of available source code, such as more niche languages like Haskell with only around 8000 active repositories \cite{GithubArchive}, these transformer models fail horrendously, sometimes unable to create runnable programs in any context. Furthermore, these transformer models can show a lack of understanding of the semantics of programs they output. Often times generated programs don't parse, don't compile, or don't match input-output examples.

We address both of these problems in this thesis. First, we make the observation that classical synthesis methods can be used to generate a large synthesized corpus of programs in any language given that language's context-free-grammar (CFG) or some other specification. We develop an algorithm that is able to, given such a grammar, able to quickly generate training examples in some given language, and even constrain generated programs to fit certain compile-time guarantees. 
Second, we note that when models are augmented with compile-time attributes or other static analysis they can perform better than they would otherwise \cite{mukherjee2021neural}. While this approach of combining static analysis with neural methods is not new, we find it to be particularly effective in low-data environments. We test the effectiveness of this approach on both a synthetic language and Haskell, utilizing static analysis of both languages in order to increase data quality in these data-poor languages.

To test these ideas, we develop a pipeline, where a corpus of random programs are synthesized alongside input/output examples and then annotated using static analysis. We then train a transformer model on the annotated synthetic corpus and evaluate the accuracy of the model. We also test the viability of static analysis on existing languages by annotating Haskell source code, feeding the annotated source into a transformer model, and then evaluating the quality of the generated source code.

We show promising results that synthesizing artificial training datasets can allow transformers to perform well on similarly synthesized tasks, even if such methods have trouble with robustness and generalization. We also show that static analysis combined with transformer methods works especially well in data-poor settings. Specifically, static analysis increases model accuracy depending on the added analysis, across both synthesizing programs given input/output examples, and for synthesizing programs given a function header. This also applies across research languages like Lambda2 and real languages like Haskell. We hope that such findings will help alleviate the widening model performance gap between high and low-data languages. 
\section{Prior and Related Work} 
\subsection{Neural-Based Program Synthesis}
Transformers are a language encoder-decoder model which namely functions off of the idea of attention. Attention is used to rank the relative importance among a set of input vectors, usually called "keys" given a single input vector known as the "query" \shortcite{transformerOrig}. These models are highly general because they work off strings as inputs, and strings as outputs, which make them theoretically generalizable to most tasks. Such tasks include synthesizing programs given natural language descriptions \shortcite{chen2021evaluating}. However, these methods often are evaluated on Python, a language with an extremely high amount of data. 

Other neural models and specifications are also frequently used for program synthesis: Neural models which function off of input-output examples but use other neural methods, such as RNNs \shortcite{robustfill} or just plain encoder-decoder models \shortcite{deepcoder} have also been explored, with some success. 
\subsection{Domain-Aware Neural Models}
Program synthesis usually requires writing a program in a formal language, which requires some amount of formal knowledge. However, neural networks aren't usually aware of this formal knowledge unless we specifically guide them to take advantage of it. Some program synthesis models specifically force a model to generate values specific to the domain language they're evaluating on, or elect to use the neural models to guide a more formal program synthesis search \shortcite{deepcoder}. 
Methods have also been explored where neural models are augmented with a static analyzer, which analyzes code both during training and during generation, and additionally feeds model with data from the static analyzer represented in some vector form \shortcite{mukherjee2021neural}. So while the concept of augmenting program synthesis with static analysis is not new, it has not to our knowledge been applied in the context of low-data environments or transformers. 

\subsection{Formal Program Synthesis from Input/Output Examples or Function Headers}
In this paper, we explore a neural framework for synthesizing a program given input output examples, or some function header. However, fully formal methods exist for such tasks. Input/output examples have been studied in the formal setting, for both simple atomic data types like numbers and strings \shortcite{numberIOSynthesis} and for recursive data types \shortcite{l2}. Sometimes, formal methods are made to be interactive, where additional outputs given inputs are requested when a program is underspecified by the current set of examples \shortcite{spreadsheetSynthesis}.
Function headers have also been explored as a possible specification for program synthesis. In Haskell, formal synthesis tools given function headers have been explored \shortcite{djinnHaskell}. These synthesis approaches have also been explored when given refinement types, which allows for richer type definitions, and therefore more control over the synthesized program  \shortcite{refinementTypeHeaderSynthesis}. 

\section{Synthesizing Training Programs}
The process of synthesizing training programs takes an attributed grammar and hopes to generate a large number of abstract syntax trees (ASTs) in that language. A na\'ive approach could randomly expand production rules of the grammar until every leaf of the tree corresponded with a terminal production rule. While this leads to \textit{syntactically} correct programs, these programs are not always \textit{semantically} correct, which leads to them being uncompilable or uninterpretable (depending on the language). An example of this is programs being generated that use undeclared variables -- these programs are correct in the context-free-grammar but not semantically correct.

In order to ensure that synthesized programs are more often semantically correct, rules in our grammar are annotated with procedures that generate attributes. Because each node corresponds with a production rule, each node has a set of attributes that is calculated using the procedure defined by the production rule. We distinguish between \textit{synthesized attributes}, which are attributes that can be determined solely from the attributes of that node's children, and \textit{inherited attributes}, which are attributes that can be determined solely from the attribute of that node's left-siblings and parent. Examples of attributes are the return type of that node in the AST, or the set of variables that are available in scope at this point in the program.

With access to these attributes, we can define constraints on the attributes that our generated programs have to follow. Constraints on using correct types and variables proves to be sufficient to generate semantically programs much more frequently.

One important property of our synthetic attributes is that they are reversible during the program generation process. For example, say we have an attribute function $f$ that simply takes the attributes of its first child. Such an attribute-generating function would look like this:
$$f(node) = f(children(node)[0])$$
However, if we have some constraint $C$ on the attributes of a node that has this attribute-generating function on it, we can solve this constraint by solving it on the first child:
$$\frac{f(node) \vdash C}{f(children(node)[0]) \vdash C}$$
In other words, we can turn constraints on a node with some synthetic attribute generating function into equivalent constraints on that node's children. During program generation, this will be at the core of our constraint solving algorithm: We can solve for constraints on a single node of the AST by solving for a different but equivalent set of constraints on its children. With this, we can make an extended program generation algorithm that takes into account attributes and constraints:

\begin{algorithm}
\caption{$generateASTNode(S, C)$: Generate a program of symbol type $S$ with constraints $C$}\label{alg:cap}
\begin{algorithmic}
\Require $S$ is the symbol in the grammar wanted
\Require $C$ is the set of constraints on attributes
\Ensure Returns a node of an AST who's attributes satisfy those constraints.
\State $A \gets$ attribute sets that satisfy constraints
\State $N \gets $ an empty AST node.
\State $P \gets$ set of all production rules with $S$ on the left-hand-side.
\State $tries \gets 0$
\While{$attrs(N)$ does not satisfy $C$ and $tries < maxTries$}
    \State $N \gets $ an empty AST node.
    \State $P' \gets randomElementOf(P)$
    \State $C' \gets equivalentConstraintsOnEachChild(P', C)$
    \State $A' \gets randomElementOf(A)$
    \For{$i=0$ to $i = len(rhs(P))$}
        \State $rhsSymbol \gets rhs(P)[i]$
        \State $child \gets generateASTNode(rhsSymbol, C'[i])$
        \State $children(N)[i] \gets child$
    \EndFor
    \State $tries \gets tries + 1$
\EndWhile
\If {$tries = maxTries$}
    \State \Return FAIL
    \Else \State \Return $N$
\EndIf
\end{algorithmic}
\end{algorithm}

It's important to note that while attributes are not strictly necessary for creating the synthetic corpus, they substantially decrease the amount of time it takes to generate a corpus while also increasing the corpus quality, and providing a large amount of auxiliary information about programs that we can pass on to the transformer in order to aid it.

In order to generate input-output examples for L2, we would analyze the AST of the L2 program in order to figure out the type of inputs to the program. For boolean inputs, we'd simply generate both true and false values. For integers, we uniformly sampled from -5 to 15. For lists, we'd generate a list of random size from 0 to 5 and generate each element randomly and independently.

\subsection{L2 Constraints and Attributes}
Recall that for our attribute grammars, we specified that every production rule could specify some function $f$ that computed the synthesized attributes: The attributes that only depend on the node's children. In order to allow the constraint solving algorithm to function quickly, these attribute-generating functions have to be able to translate constraints on a node into a set of constraints on each child. We'll go over some examples of such constraint functions now.

\subsubsection{Type Constraints}
Frequently, one wants a constraint on the type of the expression to generate. For example, say I want to generate an expression of type Int, and I'm considering using the production rule:
$$\langle Expr \rangle \to \langle Plus \rangle \langle Expr \rangle \langle Expr \rangle $$
The constraint-modifying function for this program would be simple: This production rule will always result in a node that's annotated with a type Int, provided it's given two children that are also Ints. Therefore, our attribute-generating function is super-simple: $type(f(children)) = "Int"$. Our constraint modifications can be written like this:
\begin{equation}
\frac{f(node) \vdash type(f(node)) = "Int"}{
\splitfrac{\textstyle f(children(node)[1]), f(children(node)[2]) \vdash}
{\textstyle type(f(children(node)[1])) = "Int"
\textstyle  \land type(f(children(node)[2])) = "Int"}}
\end{equation}
During dataset generation, this would result in a hole that's constrained to be an integer to result in having to generate two expressions constrained to be integers. This, while valid, isn't ideal, because we've turned one problem into two problems. Fortunately, constants exist. For example, a production rule of the form:
$$\langle Expr \rangle \to 1$$
Has the exact same attribute generating function as addition. However, it's constraint modification process is super easy:
$$\frac{f(node) \vdash type(f(node)) = "Int"}{\top}$$
In other words, if you want an integer, and you use that rule, you get an integer without any sub-goals. Easy!
\subsubsection{List and Length Constraints}
Attribute values are generic, and therefore can return integers, too. It can be useful to arbitrarily bound the length of lists that a grammar can produce. Imagine a set of rules that allow a grammar to produce a list of arbitrary size:
$$\langle List \rangle \to \langle List \rangle \langle Unit \rangle$$
$$\langle List \rangle \to \langle Unit \rangle$$
In other words, a list is either a single element, or a single element appended to another list. If you want to allow for empty lists, you could replace the right-hand-size of the second rule with some "nil" symbol.
We can create synthetic attribute functions in order to recursively calculate the length of a list. For our two rules, our attribute functions become:
$$length(f(node)) = length(f(children(node)[0])) + 1$$
$$length(f(node)) = 1$$
Likewise, we can make constraint modifiers that tell our generator how to generate lists of some arbitrary size $n$:
$$\frac{f(node), n \geq 2 \vdash length(f(node)) = n}{f(children(node)[0]) \vdash length(f(children(node)[0])) = n - 1}$$
And for our "unit" rule:
$$\frac{f(node) \vdash length(f(node)) = 1}{\top}$$
In this way, we can constrain the length of a list generated by a grammar. This is quite helpful for things like lambda declarations, which can have a variable number of inputs. If I want to constraint a lambda declaration to only have 2 inputs, then constraining the length of the list of arguments is the way to do it.

\subsubsection{Name Constraints}
Frequently, a language will have variables that can have many types. Depending on the program, the variable "x" may be an integer, a boolean, or something else entirely. It may even be undeclared! Imagine the rule:
$$\langle Var \rangle \to \backslash w+ $$

Where $Var$ can be any variable name, and $\backslash w+$ is the regex expression for a valid variable name. This will result in the attribute:
$$varname(f(node)) = stringRepresentation(children(node)[0])$$
We'd obviously want $Var$ to be a variable that's been declared, and we may even want $Var$ to have some specific type. However, knowledge of which variables have been declared in the past is not something that can be determined from the children of a node: Instead, these attributes would be \textit{inherited attributes}, because they depend on the attributes of left-siblings and parent nodes, because variable declarations happen before they are used. This turns out to not be a problem: During the program generation process, we know the inherited attributes of a node, so we can use those in the definition of our constraint modifying process: For notational purposes, when $f$ is the attribute generating function, we'll say $f_i$ is the inherited attribute function, and $f_s$ is the synthetic attribute function, such that $f = f_s \cup f_i$:
$$\frac{varname(f(node)) \vdash isDeclared(varname(f_i(node)))}{f(children(node)[0]) \vdash isDeclared(stringRepresentation(f(children(node))[0]))}$$
This will force the random string that is selected to be the variable name to be declared. This process can be repeated for forcing variables to have specific types.

\subsection{String Representations of Programs}
As transformer-based methods fundamentally function on sequences/strings, finding a string representation of our programs is necessary in order for our model to be able to process our corpus. The trivial representation is to simply turn a program AST into a string by concatenating the strings at leaf nodes: After all, terminals of an AST are all strings. However, we have access to a full parse-tree of the program, annotated with helpful type and state data! Instead of simply outputting the tokenized string representation of the program, we find that outputting a string representation of \textbf{the entire parse tree and its attributes} gives the transformer model access to deeper information about the program. This matches findings in transformer-based program execution \cite{nye2021work} and Java program synthesis \cite{mukherjee2021neural} that giving neural models access to program internal state and additional semantic information about the program improves the performance of such models.

\section{Implementation}
\subsubsection{L2 Analysis}
In order to test and evaluate this system, we chose the Lambda2 (L2) functional language outlined in \cite{l2}. This is because it almost certainly has no presence in any pre-trained model's training data, and is a relatively simple grammar with easy-to-compute attributes, while being able to express a wide range of programs by working with higher-order functions and algebraic data types. For ease of readability, we implemented the L2 language as resembling Python in syntax. Here's an example of a L2 program to find the maximum of a list of lists:
\begin{lstlisting}[
    basicstyle=\normalsize, %or \small or \footnotesize etc.
]
lambda x : max(map(lambda l : max(l), x))
\end{lstlisting}
And here's a L2 program to find the last element of a list:
\begin{lstlisting}[
    basicstyle=\normalsize, %or \small or \footnotesize etc.
]
lambda x : indexinto(x, minus(len(x), 1))
\end{lstlisting}

The process that takes program ASTs and turns them into strings creates a tab-indented tree string, where each line is a node coupled with a list of that node's attributes, and tab indents indicate child-parent relationships between nodes. Next, we assigned a unique integer ID to each production rule in our grammar, so that we wouldn't overload the transformer model with too many tokens from printing out entire production rules. Finally, terminal nodes were just represented as quoted strings. For example, the program "lambda r : neg ( r )" where $r$ is a boolean has the following AST string representation:
\begin{lstlisting}[
    basicstyle=\small, %or \small or \footnotesize etc.
]
3 {type=bool, 0.type=bool, length=1}
    "lambda"   {}
    4 {0.type=bool, length=1, r_is_decl=true, r_typevar=bool, c_is_decl}
        1 {r_is_decl=true, r_typevar=bool, type=bool}
            39 {chosenSymbol=r, type=bool}
                "r"   {}
    ":"   {}
    25 {type=bool, 0.type=bool, r_is_decl=true, r_typevar=bool}
        "neg"   {}
        "("   {}
        0 {type=bool, r_is_decl=true, r_typevar=bool}
            2 {type=bool, r_is_decl=true, r_typevar=bool}
                39 {type=bool, r_is_decl=true, r_typevar=bool}
                    "r"   {}
        ")"   {}
\end{lstlisting}
As you can see, this representation of programs allows for variable types, declared variables, and types of sub-expressions to be visible for the transformer model.

Instead of training a transformer model from scratch, we fine-tuned an existing model from GPT-Neo \cite{gpt-neo}. We chose the 125 million parameter model from GPT-Neo for two reasons: First, it was trained on source code, which likely shares patterns no matter what the language. Second, we chose it because it was what could fit on the GPUs we had access to. Further work could be done to evaluate this method given larger transformer models.

\subsection{Haskell Analysis}
In order to test the performance of our data augmentation methods on Haskell, we had to be able to actually perform static analysis on large amounts of Haskell code. Fortunately, the Glasgow Haskell Compiler (GHC), which is the biggest compiler for Haskell, exposes a number of APIs that we're able to use in order to do this. Parsing is simple: The GHC exposes an interface for parsing source code into an AST, which we can serialize in order to feed to the transformer. Nearly all source code is parsable, and even unit tests and helper scripts could be parsed, which allowed us to grab a much, much larger training dataset for parse trees.

Typechecking was more difficult. A file in source code has to be able to fully compile in order to be typechecked. This means when calling the GHC API in order to compile a package, we have to mimic the intended compile flags and compilation environment that it's intended to be compiled with. Haskell has a common build system called Cabal which attempts to standardize the build process. However, Cabal doesn't expose the GHC API, so we can't find the intermediate representations of programs that have the information we need.

We found a workaround: GHC exposes an API for plugins, which can modify and access intermediate representations of programs during compilation. We created a plugin that would take those representations, annotate them with types, and serialize them to a file. Then, by hardcoding our installation of GHC to use that plugin, every package that can be compiled with Cabal uses that plugin. This still didn't compile helper scripts and unit tests, and not all packages could be compiled this way, so it resulted in a much smaller dataset. In total, we ended up with 16,000 type-annotated functions, and 40,000 parsed functions.

Serializing these Haskell trees into a consistent string format was non-trivial. Writing serialization functions for every possible node type in the GHC would've taken far too much time, and would've likely broken upon the release of any new GHC version. Many data types in GHC already had ways of serializing and deserializing. However, there were many types for which there was no easily definable generic function. In order to fix this, we had to write several special cases for types in the GHC which did not support serialization. When doing this, the serialization and deserialization methods were vert lossy: Much data was thrown away in the interest of the compactness of the string representations. However, we made sure that we would not throw away data that was necessary for recreating the original source code given it's tree representation. For example, consider the Var type in GHC, the datatype which represents a variable. This was a type that had special constructors that didn't allow for generic serialization and deserialization. In order to create a Var normally, you'd have to choose the right helper function to make it, which would usually require the variable's type, all subtypes and possible algebraic types resulting from that type, knowledge of the variable's scope, knowledge of the module that variable was declared in, and so on. However, we just store the string representation of the variable, and leave placeholder values for everything else. After all, when turning the the AST string into a source string, we only need the variable's name. 

Therefore, if you were to take source code, serialize it's tree, and then deserialize it's tree, you would lose much information. But, if you take  code, serialize it's tree, and then deserialize it's tree, and turn that back into source code, you get back source that's nearly identical. This was good enough for our purposes. 
\section{Evaluation}
\subsection{Evaluation on synthesized L2}
To evaluate on L2, a synthesized language, we generate 10,000 training examples and 1,000 evaluation examples. These examples have 4 to 7 input/output pairs each. In order to not end up with an adversely simple training dataset, the training dataset is pruned of programs that only return a constant, or programs that are equivalent to the identity function. 

To evaluate our model, we generate 1000 programs each with a set of 4-7 input/output examples. We then mask the programs out, and have GPT generate a program when given the examples.

We evaluated multiple models: Models trained on the raw string representation of the program, models trained on the program's AST with and without attributes, and models trained on program ASTs bundled with a post-processing Sketch algorithm, which post-processed generated programs by masking out constants and variables and then doing an enumerative search over possible constants to "fill" the masked holes in the program. This was done after noticing that there were a high amount of programs that would be correct structurally, but GPT would make minor errors such as multiplying by 3 instead of 2, or other constant-related issues.  

We also found that when generating programs for L2, some symbols and functions appeared far more often in incorrect programs than correct ones. The terminal symbols that were most common in incorrectly generated programs but not in correctly generated programs, alongside the difference in frequencies between correct and incorrect programs, were:
\begin{enumerate}
    \item times (2.5\%)
    \item minus (2.3\%)
    \item plus (2.0\%)
    \item lt (0.9\%)
    \item min (0.2\%)
    \item gt (0.2\%)
\end{enumerate}
Namely, mostly arithmetic and integer operations. So, we did another run of the model but trained and evaluated on a subset of L2 that only contained list operations and higher-order operators on lists.

Next, another experiment that we did was to increase the number of programs generated by the transformer model. In hindsight, it may be unreasonable to expect a transformer model to generate a fully correct program on the first try. Furthermore, generating programs with the transformer and verifying the correctness of a program against input/output examples is computationally cheap. Therefore, we had GPT generate multiple programs per set of input/output examples and then selected the one that satisfied the most examples. 

Finally, we also evaluated our best performing model on a series of handmade programs that perform basic arithmetic and list manipulation. This dataset included classical algorithms such as reversing or finding the length of lists, removing even elements from list, finding the cartesian product of a list of lists, and so on. We used the same programming problems that L2 was originally evaluated on \shortcite{l2} that involved list processing or arithmetic. We did this to see if the model, when trained on synthetic programs and problems, can generalize to other, more "real" program synthesis tasks. 

\begin{table*}[t]
  \centering
  \begin{tabular}{|| c|| c | c | c | c | c | c | c || }
  \hline
 Method & Correct & Wrong Output & Parse & Type & Decode & Name & Runtime \\
    \hline
    \multicolumn{8}{|| c ||}{Evaluated on Synthesized Data} \\
    \hline
    Trees+Attributes+NoArith+25 & 5431 & 375 & 0 & 0 & 2 & 0 & 31 \\
    Trees+Attributes+25Tries & 3798 & 2105 & 0 & 0 & 12 &t 0 & 57 \\
    Trees+Attributes+10Tries & 3420 & 2449 & 0 & 0 & 13 & 0 & 90 \\
    Trees+Attributes+Sketch & 2667 & 3218 & 0 & 0 & 0 & 0 & 138 \\
    Trees+Attributes & 2637 & 3237 & 0 & 0 & 7 & 0 & 142 \\
    Trees+No Attrs & 2366 & 3269 & 0 & 124 & 0 & 46 & 126 \\
    Plain Program & 2111 & 2555 & 23 & 211 & 0 & 10 & 83 \\
    \hline
    \multicolumn{8}{|| c ||}{Evaluated on Handmade Data} \\
    \hline 
    Trees+Attributes+25Tries & 34 & 66 & 0 & 0 & 4 & 0 & 15 \\
    \hline
  \end{tabular}
  \caption{Types of errors encountered in GPT-generated programs}
  \label{tab:1}
\end{table*}

With the knowledge that adding attributes and allowing for multiple "attempts" at a correct program tends to increase results, we also experimented with increasing the size of the training data by synthesizing more programs. 
\begin{table*}[t]
  \centering
  \begin{tabular}{|| c || c | c | c | c | c | c | c || }
  \hline
 Train \# & Correct & Wrong Output & Parse & Type & Decode & Name & Runtime \\
    \hline
    \multicolumn{8}{|| c ||}{Evaluated on Synthesized Data} \\
    \hline
    1k & 3085 & 1896 & 0 & 0 & 292 & 0 & 692 \\
    10k & 3965 & 1810 & 0 & 0 & 42 & 0 & 157 \\
    100k & 4422 & 1410 & 0 & 0 & 48 & 0 & 92 \\
    1M & 4921 & 787 & 0 & 0 & 220 & 0 & 44 \\
    \hline
    \multicolumn{8}{|| c ||}{Evaluated on Handmade Data} \\
    \hline
    1M & 42 & 60 & 0 & 0 & 10 & 0 & 7 \\
    \hline 
    
    \hline
  \end{tabular}
  \caption{Types of errors encountered in GPT-generated programs by training size}
  \label{tab:1}
\end{table*}

We distinguish between examples that have the following errors: 
\begin{enumerate}
    \item Parse errors are examples that fail because the programs that run them aren't can't be parsed, or it's impossible to create syntax trees from them. Note that it's impossible to get parse errors in the cases where we generate full ASTs. 
    \item Type errors are examples that fail because of a type mismatch, such as attempting to find the length of a boolean. 
    \item Decode errors are examples that fail because the transformer model has generated a string that's not a valid abstract syntax tree, in the cases where we generate a tree and not the program itself. Note that it's impossible to get decode errors unless we're generating an AST. 
    \item Name errors are errors caused by referencing undeclared variables or re-declaring existing variables
    \item Runtime errors are just examples that are correct but fail at runtime. This can be like dividing by zero, or trying to find the minimum element of an empty list, or indexing an array out of bounds, or anything else.  
\end{enumerate}
\subsection{Evaluation on Haskell}
For Haskell, we had three corpuses we were able to train on. One was a corpus of plain Haskell source code, one was a corpus of Haskell functions turned into an AST representation, and one was a corpus of Haskell functions with AST string representations with type-annotated nodes. In order to evaluate these systems, we generated each data format's equivalent to a function header that maps integers to integers. As an example, this is represented in plain Haskell as "f :: Int -> Int". This wold be the prompt to our model. We then generated a thousand programs with this as the prompt. Because there are no input/output examples for this task, we simply counted a program as correct if it successfully ran on an integer: we tested 0 and 1. The evaluation and error types were otherwise the same as when evaluating L2.  
\begin{table*}[t]
  \centering
  \begin{tabular}{|| c || c | c | c | c | c | c || }
  \hline
 Method & Valid Output & Parse & Type & Decode & Name & Runtime \\
    \hline
    Plain Haskell & 0 & 978 & 6 & 0 & 0 & 16 \\
    Parsetree & 26 & 2 & 40 & 821 & 96 & 15 \\
    Typed Parsetree & 10 & 15 & 30 & 464 & 478 & 3 \\
    \hline
    \hline
    
    \hline
  \end{tabular}
  \caption{Types of errors encountered in GPT-generated programs on Haskell}
  \label{tab:3}
\end{table*}

\section{Conclusions}
From our results, we draw a few conclusions. Firstly, datasets for low-data languages can be efficiently synthesized using formal methods, to some success. However, we also found that models trained on synthetic program corpuses tend to not generalize well to other kinds of programs or datasets. This matches other findings with other ML-based program synthesis techniques similar, in which models can get incredibly high accuracy on synthetic datasets but perform poorly on handmade ones \cite{suh2021adversarial}. Theoretically, if our synthesized corpus more closely matched the kinds of programs we'd expect a human would want to make, or if our synthesized corpus was more diverse and therefore trained a more robust model, this performance difference might disappear.

Second, we conclude that, especially in the low-data case, giving transformer models access to additional information about programs enables better program generation. We saw type and name errors completely eliminated when feeding type and variable scoping attributes to the model. Second, we found that the model performs especially poorly on arithmetic -- this makes sense, as we suspect the embeddings/vector representations for integers tend to be very close and nearly indistinguishable from each other, so choosing the correct integer for programs that involve complex arithmetic is tricky. 

Finally, we've also shown the capability of alternative representations of programs in order to improve performance. For real low-data languages like Haskell, we've shown that simple data augmentation using static analysis is able to greatly improve the quality of generated programs. 

We hope that these methods help bridge the data gap between languages, and hopefully enable models to perform better synthesis regardless of the popularity of the chosen programming language.

\subsection{Future Work}
Much room for improvement exists in these methods. Firstly, tuning the distribution of the training dataset away from our uniform sampling is possible to do. Theoretically, one could use a probabilistic grammar or some variant to tune the distribution of the training set to create more "human-like" programs. However, defining what makes a program "human-like" is tricky. Finding a distribution would result in "human-like" programs is similarly difficult. Theoretically, one could use generative adversarial networks or some other neural method in order to learn a distribution that performs well, but we did not explore this pathway. 

Furthermore, more work exists in synthesizing programs in real languages. The grammar for real languages like Haskell is incredibly complete, even when ignoring attributes and constraints. Formulating any non-trivial language in the method we describe poses a non-trivial engineering challenge. However, a plus side for working in real languages would be that we can draw a distribution of training programs and their ASTs from their small amount of existing source code, and hopefully use that to solve the problem of training programs not being "human-like". 

\section*{Appendix}

\subsection{Our representation of the L2 Language}

The L2 Language, as we tested it, contains two atomic types (Ints and Bools), one algebraic List type, and a function type. We implemented most basic boolean functions, and the basic arithmetic operations of addition, subtraction, multiplication, division, remainder, less than, and greater than. We also implemented basic list operations, such as indexing, concatenation, list insertion, minimum, maximum, and list length. Finally, we also added higher-order functions, such as map, filter, reduce, foldl, foldr, and recl.

The grammar for L2 is extraordinarily simple: An L2 Program is an expression that is some function type. Each expression can be a function declaration, a constant, or a library function, where the arguments to each library function are themselves L2 expressions.

\subsection{Source Code}

The full library and pipeline for generating a training set, training GPT, using GPT, and then evaluating the results can be found at this repo: \url{https://github.com/JRoper18/cfgenerator}

The library and accompanying plugin for GHC that parses, typechecks, annotates, and serializes Haskell source code can be found here: \url{https://github.com/JRoper18/haskell-scraper}

\bibliography{main}

\begin{thebibliography}{16}
\expandafter\ifx\csname natexlab\endcsname\relax\def\natexlab#1{#1}\fi

\bibitem[{Balog et~al.(2016)Balog, Gaunt, Brockschmidt, Nowozin, and
  Tarlow}]{deepcoder}
Matej Balog, Alexander~L. Gaunt, Marc Brockschmidt, Sebastian Nowozin, and
  Daniel Tarlow. 2016.
\newblock \href {https://doi.org/10.48550/ARXIV.1611.01989} {Deepcoder:
  Learning to write programs}.

\bibitem[{Barrett et~al.(2011)Barrett, Conway, Deters, Hadarean, Jovanovic,
  King, Reynolds, and Tinelli}]{DBLP:conf/cav/BarrettCDHJKRT11}
Clark~W. Barrett, Christopher~L. Conway, Morgan Deters, Liana Hadarean, Dejan
  Jovanovic, Tim King, Andrew Reynolds, and Cesare Tinelli. 2011.
\newblock \href {https://doi.org/10.1007/978-3-642-22110-1\_14} {{CVC4}}.
\newblock In \emph{Computer Aided Verification - 23rd International Conference,
  {CAV} 2011, Snowbird, UT, USA, July 14-20, 2011. Proceedings}, volume 6806 of
  \emph{Lecture Notes in Computer Science}, pages 171--177. Springer.

\bibitem[{Black et~al.(2021)Black, Gao, Wang, Leahy, and Biderman}]{gpt-neo}
Sid Black, Leo Gao, Phil Wang, Connor Leahy, and Stella Biderman. 2021.
\newblock \href {https://doi.org/10.5281/zenodo.5297715} {{GPT-Neo: Large Scale
  Autoregressive Language Modeling with Mesh-Tensorflow}}.
\newblock {If you use this software, please cite it using these metadata.}

\bibitem[{Chen et~al.(2021)Chen, Tworek, Jun, Yuan, de~Oliveira~Pinto, Kaplan,
  Edwards, Burda, Joseph, Brockman, Ray, Puri, Krueger, Petrov, Khlaaf, Sastry,
  Mishkin, Chan, Gray, Ryder, Pavlov, Power, Kaiser, Bavarian, Winter, Tillet,
  Such, Cummings, Plappert, Chantzis, Barnes, Herbert-Voss, Guss, Nichol,
  Paino, Tezak, Tang, Babuschkin, Balaji, Jain, Saunders, Hesse, Carr, Leike,
  Achiam, Misra, Morikawa, Radford, Knight, Brundage, Murati, Mayer, Welinder,
  McGrew, Amodei, McCandlish, Sutskever, and Zaremba}]{chen2021evaluating}
Mark Chen, Jerry Tworek, Heewoo Jun, Qiming Yuan, Henrique~Ponde
  de~Oliveira~Pinto, Jared Kaplan, Harri Edwards, Yuri Burda, Nicholas Joseph,
  Greg Brockman, Alex Ray, Raul Puri, Gretchen Krueger, Michael Petrov, Heidy
  Khlaaf, Girish Sastry, Pamela Mishkin, Brooke Chan, Scott Gray, Nick Ryder,
  Mikhail Pavlov, Alethea Power, Lukasz Kaiser, Mohammad Bavarian, Clemens
  Winter, Philippe Tillet, Felipe~Petroski Such, Dave Cummings, Matthias
  Plappert, Fotios Chantzis, Elizabeth Barnes, Ariel Herbert-Voss,
  William~Hebgen Guss, Alex Nichol, Alex Paino, Nikolas Tezak, Jie Tang, Igor
  Babuschkin, Suchir Balaji, Shantanu Jain, William Saunders, Christopher
  Hesse, Andrew~N. Carr, Jan Leike, Josh Achiam, Vedant Misra, Evan Morikawa,
  Alec Radford, Matthew Knight, Miles Brundage, Mira Murati, Katie Mayer, Peter
  Welinder, Bob McGrew, Dario Amodei, Sam McCandlish, Ilya Sutskever, and
  Wojciech Zaremba. 2021.
\newblock \href {http://arxiv.org/abs/2107.03374} {Evaluating large language
  models trained on code}.

\bibitem[{Devlin et~al.(2017)Devlin, Uesato, Bhupatiraju, Singh, Mohamed, and
  Kohli}]{robustfill}
Jacob Devlin, Jonathan Uesato, Surya Bhupatiraju, Rishabh Singh, Abdel-rahman
  Mohamed, and Pushmeet Kohli. 2017.
\newblock \href {https://doi.org/10.48550/ARXIV.1703.07469} {Robustfill: Neural
  program learning under noisy i/o}.

\bibitem[{Feser et~al.(2015)Feser, Chaudhuri, and Dillig}]{l2}
John~K. Feser, Swarat Chaudhuri, and Isil Dillig. 2015.
\newblock \href {https://doi.org/10.1145/2813885.2737977} {Synthesizing data
  structure transformations from input-output examples}.
\newblock \emph{SIGPLAN Not.}, 50(6):229–239.

\bibitem[{Gao et~al.(2021)Gao, Biderman, Black, Golding, Hoppe, Foster, Phang,
  He, Thite, Nabeshima, Presser, and Leahy}]{thepile}
Leo Gao, Stella Biderman, Sid Black, Laurence Golding, Travis Hoppe, Charles
  Foster, Jason Phang, Horace He, Anish Thite, Noa Nabeshima, Shawn Presser,
  and Connor Leahy. 2021.
\newblock \href {https://doi.org/10.48550/ARXIV.2101.00027} {The pile: An 800gb
  dataset of diverse text for language modeling}.

\bibitem[{Grigorik(2008)}]{djinnHaskell}
Ilya Grigorik. 2008.
\newblock Djinn: Generate haskell code from a type.
\newblock \url{https://hackage.haskell.org/package/djinn}.

\bibitem[{Grigorik(2022)}]{GithubArchive}
Ilya Grigorik. 2022.
\newblock Github archive.
\newblock \url{https://github.com/igrigorik/gharchive.org}.

\bibitem[{Gulwani(2011)}]{spreadsheetSynthesis}
Sumit Gulwani. 2011.
\newblock \href {https://doi.org/10.1145/1926385.1926423} {Automating string
  processing in spreadsheets using input-output examples}.
\newblock volume~46, pages 317--330.

\bibitem[{Mukherjee et~al.(2021)Mukherjee, Wen, Chaudhari, Reps, Chaudhuri, and
  Jermaine}]{mukherjee2021neural}
Rohan Mukherjee, Yeming Wen, Dipak Chaudhari, Thomas~W. Reps, Swarat Chaudhuri,
  and Chris Jermaine. 2021.
\newblock \href {http://arxiv.org/abs/2111.01633} {Neural program generation
  modulo static analysis}.

\bibitem[{Nye et~al.(2021)Nye, Andreassen, Gur-Ari, Michalewski, Austin,
  Bieber, Dohan, Lewkowycz, Bosma, Luan, Sutton, and Odena}]{nye2021work}
Maxwell Nye, Anders~Johan Andreassen, Guy Gur-Ari, Henryk Michalewski, Jacob
  Austin, David Bieber, David Dohan, Aitor Lewkowycz, Maarten Bosma, David
  Luan, Charles Sutton, and Augustus Odena. 2021.
\newblock \href {http://arxiv.org/abs/2112.00114} {Show your work: Scratchpads
  for intermediate computation with language models}.

\bibitem[{Polikarpova et~al.(2016)Polikarpova, Kuraj, and
  Solar-Lezama}]{refinementTypeHeaderSynthesis}
Nadia Polikarpova, Ivan Kuraj, and Armando Solar-Lezama. 2016.
\newblock \href {https://doi.org/10.1145/2980983.2908093} {Program synthesis
  from polymorphic refinement types}.
\newblock \emph{SIGPLAN Not.}, 51(6):522–538.

\bibitem[{Singh and Gulwani(2012)}]{numberIOSynthesis}
Rishabh Singh and Sumit Gulwani. 2012.
\newblock Synthesizing number transformations from input-output examples.
\newblock In \emph{Computer Aided Verification}, pages 634--651, Berlin,
  Heidelberg. Springer Berlin Heidelberg.

\bibitem[{Suh and Timen(2021)}]{suh2021adversarial}
Alexander Suh and Yuval Timen. 2021.
\newblock \href {https://openreview.net/forum?id=aI8VuzSvCPn} {Adversarial
  synthetic datasets for neural program synthesis}.

\bibitem[{Vaswani et~al.(2017)Vaswani, Shazeer, Parmar, Uszkoreit, Jones,
  Gomez, Kaiser, and Polosukhin}]{transformerOrig}
Ashish Vaswani, Noam Shazeer, Niki Parmar, Jakob Uszkoreit, Llion Jones,
  Aidan~N. Gomez, Lukasz Kaiser, and Illia Polosukhin. 2017.
\newblock \href {https://doi.org/10.48550/ARXIV.1706.03762} {Attention is all
  you need}.

\end{thebibliography}
\bibliographystyle{acl_natbib}

\appendix



\end{document}